\documentclass[aps,prb,reprint,superscriptaddress]{revtex4-2}

\usepackage{graphicx}
\graphicspath{{figures/}}
\usepackage{amsmath,amssymb,bm}
\usepackage[colorlinks, linkcolor=blue, citecolor=blue, urlcolor=blue]{hyperref}
\usepackage{siunitx}

\begin{document}

\title{A Fully Ab-Initio Spin-Lattice Dynamics Framework for Magnetic Materials}

\author{Xianxi Zhang}
\affiliation{Key Laboratory of Computational Physical Sciences (Ministry of Education),
  Institute of Computational Physical Sciences, State Key Laboratory of Surface Physics,
  and Department of Physics, Fudan University, Shanghai 200433, China}

\author{Hongyu Yu}
\affiliation{Key Laboratory of Computational Physical Sciences (Ministry of Education),
  Institute of Computational Physical Sciences, State Key Laboratory of Surface Physics,
  and Department of Physics, Fudan University, Shanghai 200433, China}

\author{Liangliang Hong}
\affiliation{Key Laboratory of Computational Physical Sciences (Ministry of Education),
  Institute of Computational Physical Sciences, State Key Laboratory of Surface Physics,
  and Department of Physics, Fudan University, Shanghai 200433, China}

\author{Hongjun Xiang}
\email{hxiang@fudan.edu.cn}
\affiliation{Key Laboratory of Computational Physical Sciences (Ministry of Education),
  Institute of Computational Physical Sciences, State Key Laboratory of Surface Physics,
  and Department of Physics, Fudan University, Shanghai 200433, China}

\date{\today}

\begin{abstract}
Coupled spin-lattice dynamics (SLD) underlie a wide range of magnetic phenomena, yet a unified
first-principles framework that propagates both degrees of freedom without empirical
parameterization has remained elusive.
We present a fully \textit{ab initio} SLD approach integrated into VASP, in which interatomic
forces and effective magnetic fields are obtained at each time step from self-consistent
constrained-moment density-functional calculations.
The method is validated on four materials spanning ferromagnetic, non-collinear, and
geometrically frustrated orders, recovering the correct magnetic ground state in every case
from random initial conditions.
SLD trajectories also provide physically correlated training data for magnetic
machine-learning potentials, as demonstrated for BiFeO$_3$ by a reduction of
up to approximately one order of magnitude in energy MAE over training on
randomized spin configurations.
This framework opens a practical first-principles route to finite-temperature spin-lattice
coupled phenomena in magnetic materials.
\end{abstract}

\maketitle

\section{\label{sec:intro}Introduction}

The interplay between spin and lattice degrees of freedom plays a central role in a wide range
of condensed-matter phenomena, including ultrafast demagnetization~\cite{Beaurepaire1996,Weissenhofer2024},
magnetostructural phase transitions~\cite{Solovyev2012}, and magnon--phonon
scattering~\cite{Ruckriegel2014,Le2025}.
From a technological perspective, the development of spintronic
devices~\cite{Wolf2001,Zutic2004} and ultrafast magnetic recording~\cite{Kimel2005,Stanciu2007}
has further underscored the need for a microscopic, finite-temperature description of the
coupled evolution of spin and lattice degrees of freedom.
At the microscopic level, this coupling has two prominent contributions: the dependence of
magnetic exchange interactions $J_{ij}$ on interatomic distances, and spin-orbit-mediated
transfer of angular momentum between the magnetic and lattice subsystems.
A comprehensive understanding of these phenomena therefore requires a theoretical description
that treats lattice and spin degrees of freedom on an equal footing.
While static first-principles calculations have successfully elucidated the ground-state
properties of magnetic materials~\cite{Hohenberg1964,KohnSham1965}, describing
finite-temperature thermodynamics and nonequilibrium evolution necessitates a dynamical
approach that explicitly accounts for the coupling between spins and the lattice.

Conventionally, spin dynamics and their coupling to the lattice are simulated by constructing
model Hamiltonians with parametrized exchange couplings $J_{ij}$ from first-principles
calculations and propagating the spins via the Landau--Lifshitz--Gilbert (LLG)
equation~\cite{Antropov1996,Skubic2008,Evans2014,Hellsvik2019,Stockem2018,Gambino2023}.
Although computationally efficient, these methods are limited by the fixed nature of the
parametrized couplings, which cannot capture regimes where exchange interactions are strongly
modulated by lattice distortions or where the electronic structure changes significantly from the
reference state used in the fitting.
In contrast, standard \textit{ab initio} molecular dynamics (AIMD) provides an accurate
first-principles description of lattice dynamics, but it does not treat spin orientations as
independent dynamical variables and therefore neglects the transverse spin fluctuations and
precessional dynamics central to LLG evolution.
Time-dependent density-functional theory can in principle provide a fully first-principles
description of coupled electronic, spin, and lattice dynamics, but its computational cost
currently limits its use for the length and time scales relevant to finite-temperature
simulations in realistic materials~\cite{Runge1984,Ullrich2012}.
What remains lacking, therefore, is a practical first-principles framework in which the
effective magnetic fields are obtained directly from DFT at each time step, eliminating the
intermediate mapping onto parametrized exchange interactions.

In this work, we present a fully \textit{ab initio} SLD method seamlessly integrated into the
Vienna \textit{Ab initio} Simulation Package (VASP)~\cite{Kresse1996a,Kresse1996b}.
Our approach couples the Newtonian equations of motion for the lattice with the stochastic LLG
equation for the magnetic moments.
Crucially, both the interatomic forces and the effective magnetic fields that drive the spin
dynamics are obtained self-consistently at every time step from constrained local-moment DFT
calculations, without requiring an intermediate mapping onto parametrized exchange
interactions~\cite{Ma2015}.
This framework therefore enables the simultaneous evolution of atomic and spin degrees of
freedom on the same first-principles footing, allowing the dynamic coupling and energy exchange
between the lattice and spin subsystems to be captured directly.
We validate the implementation across four representative systems spanning distinct electronic
and magnetic regimes---a two-dimensional semiconductor (CrI$_3$), an itinerant ferromagnet
(bcc~Fe), a correlated 5$d$ oxide (Cd$_2$Os$_2$O$_7$), and a geometrically frustrated spinel
(MgCr$_2$O$_4$)---and find that, in all four cases, the simulations evolve from random initial
conditions to the correct magnetic ground state without ad hoc parameter adjustments.

Beyond enabling direct spin-lattice simulations, the present \textit{ab initio} SLD
framework also provides a route to generating training data for spin-aware
machine-learning models.
Direct first-principles SLD nevertheless remains computationally demanding for
larger supercells, longer trajectories, and the extensive sampling needed for
finite-temperature studies.
In this context, spin-aware machine-learning models provide a natural way to
extend spin-lattice simulations to larger scales, but their performance depends
critically on the distribution and quality of the training data.
In non-magnetic systems, practical strategies for constructing training
distributions tailored to the target thermodynamic or dynamical regime are
already well developed~\cite{Behler2016,Deringer2021,Podryabinkin2017,Schran2020},
whereas for magnetic systems the choice of a suitable distribution in the
combined space of atomic and spin degrees of freedom is less straightforward.
A common baseline is to combine structural snapshots with independently
randomized magnetic configurations, but this construction does not preserve the
correlations between local lattice distortions and spin configurations, whereas
SLD trajectories retain these correlations because atomic positions and magnetic
moments co-evolve under forces and effective fields obtained from the same
first-principles calculations.
This motivates a controlled benchmark for an 80-atom BiFeO$_3$ supercell, in
which we compare SpinGNN, a spin-aware graph neural network model, trained on
SLD-derived data with the same model trained on a reference dataset constructed
from the same initial structures but paired with randomized magnetic
configurations.
Our goal is not to claim the universal superiority of SLD-derived data, but to
test whether they offer an advantage for the physically relevant spin-lattice
configurations encountered in subsequent dynamical simulations.

\section{\label{sec:method}Methodology and Implementation}

\subsection{Theoretical Framework}

The present formulation employs two approximations commonly adopted in first-principles
spin-dynamics studies.
The first is the \emph{spin-adiabatic approximation}, which assumes that the orientations of
local magnetic moments vary on a much slower time scale than their magnitudes, so that the moment
magnitudes adapt instantaneously to the local environment and do not require their own equations
of motion, with the orientations being treated as classical dynamical variables~\cite{Antropov1996}.
The second is the \emph{atomic-moment approximation}, under which each atomic site $i$ carries
a local magnetic moment $\mathbf{M}_i$, defined as the spatial integral of the spin-density
distribution in the region surrounding atom $i$~\cite{Eriksson2017}.

Within these approximations, the coupled spin-lattice dynamics are governed by the classical
equations of motion for the ions and the stochastic Landau-Lifshitz-Gilbert (LLG) equation
for the spins:
\begin{align}
  m_i \ddot{\mathbf{R}}_i &= \mathbf{F}_i , \label{eq:newton}\\
  \begin{split}
    \frac{d\mathbf{M}_i}{dt} &= -\gamma\,\mathbf{M}_i \times (\mathbf{B}_i + \mathbf{b}_i) \\
      &\quad - \frac{\alpha}{|\mathbf{M}_i|}\,\mathbf{M}_i \times
        [\mathbf{M}_i \times (\mathbf{B}_i + \mathbf{b}_i)] .
  \end{split} \label{eq:LLG}
\end{align}
Here $m_i$ and $\mathbf{R}_i$ denote the mass and position of atom $i$, and $\mathbf{F}_i$ is
the force acting on that atom.
$\mathbf{M}_i$ represents the local magnetic moment on site $i$, $\mathbf{B}_i$ is the
effective magnetic field conjugate to it, and $\mathbf{b}_i$ is a stochastic field representing
thermal fluctuations in the spin subsystem.
The constants $\gamma$ and $\alpha$ are the gyromagnetic ratio and Gilbert damping parameter,
respectively.

To evaluate the effective magnetic field $\mathbf{B}_i$ required by the LLG equation, we
employ VASP's built-in constrained local-moment approach~\cite{Ma2015}.
In this scheme, a penalty term is added to the Kohn-Sham energy functional to constrain the
orientation of each local moment $\mathbf{M}_i$ towards a prescribed direction
$\hat{\mathbf{e}}_i^0$:
\begin{equation}
  E_p = \lambda \sum_i
    \left[\mathbf{M}_i
      - \hat{\mathbf{e}}_i^0
        \left(\hat{\mathbf{e}}_i^0 \cdot \mathbf{M}_i\right)\right]^2 ,
  \label{eq:penalty}
\end{equation}
Here $\lambda$ is a penalty strength parameter that enforces the constraint; in the limit
$\lambda\to\infty$ the moment orientation $\hat{\mathbf{M}}_i$ coincides exactly with
$\hat{\mathbf{e}}_i^0$.
The functional derivative of the penalty term with respect to the spin density yields a
constraining magnetic field,
\begin{equation}
  \mathbf{B}_i^{\mathrm{con}}
    = \frac{2\lambda}{M_i}
      \left[\mathbf{M}_i
        - \hat{\mathbf{e}}_i^0\,
          \left(\hat{\mathbf{e}}_i^0 \cdot \mathbf{M}_i\right)\right] ,
  \label{eq:Bcon}
\end{equation}
which by construction lies nearly perpendicular to $\mathbf{M}_i$.

At self-consistency, the total energy $E = E_{\mathrm{KS}} + E_p$ is stationary with respect
to the spin density, which directly implies
$\mathbf{B}_i^{\mathrm{eff}} = -\mathbf{B}_i^{\mathrm{con}}$, i.e., the constraining field
output by VASP equals the negative of the true effective field driving the spin
dynamics~\cite{Brannvall2026,Ma2015}.
This identity has been established analytically by Br\"{a}nnvall et al.\ and holds for any
exchange-correlation functional, independent of the mean-field structure of the
Hamiltonian~\cite{Brannvall2026}.
It has also been verified numerically via finite-difference calculations of the energy gradient
with respect to moment rotation; see Ref.~\cite{Cai2023} and the Supplemental Material for
independent confirmations.
The effective field $\mathbf{B}_i = -(1/M_i)\,\partial E_0/\partial\theta_i$ is therefore
directly available from the VASP output at each SLD step without any additional computation.

\subsection{Numerical Implementation}

The ionic equations of motion~\eqref{eq:newton} are integrated using the standard
velocity-Verlet scheme~\cite{Swope1982} implemented in VASP, while the LLG
equation~\eqref{eq:LLG} is integrated using the semi-implicit SIB algorithm of
Mentink et al.~\cite{Mentink2010}, which conserves the spin length and is numerically stable.
Convergence tests with time steps ranging from 0.2 to 2~fs showed that the spin integration
remained accurate up to 1~fs, whereas steps of $\approx 2$~fs introduced noticeable numerical
artifacts.
Unless otherwise noted, a time step of 1~fs is adopted throughout this work.

The stochastic field $\mathbf{b}_i$ in Eq.~\eqref{eq:LLG} is modeled as Gaussian white
noise interpreted in the Stratonovich sense~\cite{Mentink2010}, with zero mean and
correlations given by the fluctuation--dissipation theorem:
\begin{equation}
  \langle b_i^\mu(t)\,b_j^\nu(t')\rangle
    = \frac{2\alpha\,k_{\mathrm{B}}T}{(1+\alpha^2)\,\gamma\,M_i}\,
      \delta_{ij}\,\delta_{\mu\nu}\,\delta(t-t') ,
  \label{eq:noise}
\end{equation}
where $\mu,\nu$ label Cartesian components and $T$ is the temperature.
This relation ensures that the spin subsystem samples the canonical equilibrium
distribution at temperature $T$~\cite{Mentink2010}.

\begin{figure}[htbp]
  \centering
  \includegraphics[width=\columnwidth]{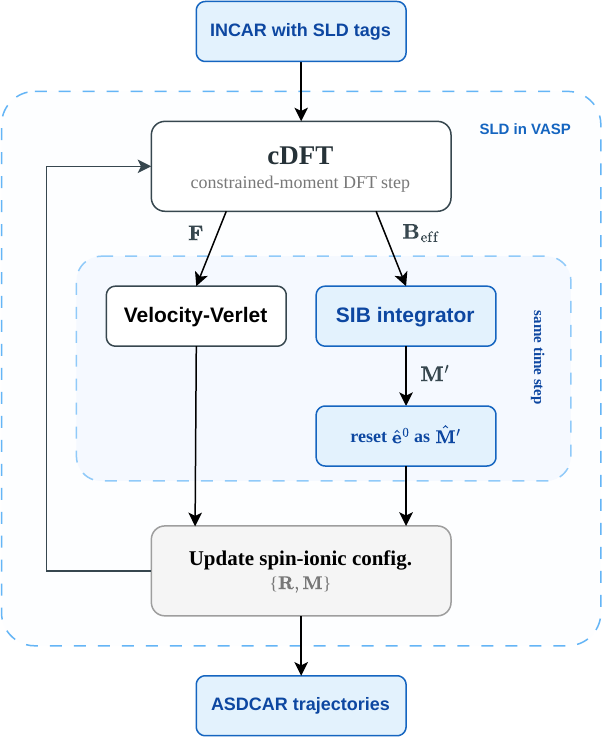}
  \caption{Schematic of the \textit{ab initio} SLD dynamical loop implemented in VASP.
           At each time step, a constrained-moment DFT calculation yields both the
           interatomic forces $\{\mathbf{F}_i\}$ and the effective magnetic fields
           $\{\mathbf{B}_i\}$, which are used to advance ionic positions via
           velocity-Verlet and spin directions via the SIB integrator, respectively.}
  \label{fig:sld_loop}
\end{figure}

As illustrated in Fig.~\ref{fig:sld_loop}, at each time step a self-consistent
constrained-moment DFT calculation yields the interatomic forces
$\{\mathbf{F}_i\}$ and the effective magnetic fields $\{\mathbf{B}_i\}$, which
then advance the ionic positions via velocity-Verlet and the spin directions via
the SIB integrator, respectively.
The updated ionic positions and spin orientations then enter the next
constrained-moment DFT calculation, with the latter defining the new constraint
directions $\hat{\mathbf{e}}_i^0$, thereby closing the dynamical loop.

This formulation is implemented within VASP, allowing direct use of its native
functionalities---including the projector-augmented-wave (PAW)
method~\cite{Blochl1994,Kresse1999}, spin-orbit coupling, and external-field
perturbations---for self-consistent spin-lattice dynamics simulations.
New tags have been introduced into the VASP input file (\texttt{INCAR}) to control the SLD
workflow, while the relevant output quantities are written to a dedicated file
(\texttt{ASDCAR}), leaving the standard VASP output unchanged.
Details of these input tags and the auxiliary analysis toolkit (\texttt{sldkit})
for post-processing and visualization are described in the Supplemental Material.

\section{\label{sec:validation}Validation via Magnetic Ground-State Reproduction}

To validate the physical accuracy of the \textit{ab initio} SLD approach, we
examine its ability to reproduce the magnetic ground states of four
representative materials with distinct electronic and magnetic characteristics.
These materials are selected because their magnetic ground states are
experimentally well established and because such magnetic orders cannot be
stabilized by standard static DFT calculations without external constraints,
thereby providing stringent benchmarks for the present approach.
Unless otherwise specified, all SLD simulations were performed in the
isothermal-isobaric ($NpT$) ensemble using a Langevin thermostat and the
Parrinello--Rahman barostat~\cite{ParrinelloRahman1981}, so that atomic
positions, lattice parameters, and magnetic moment orientations relax
simultaneously within a unified dynamical framework.
In all simulations, the Perdew-Burke-Ernzerhof (PBE) exchange-correlation
functional~\cite{Perdew1996} was employed. In addition, a relatively large Gilbert
damping parameter was adopted
solely to accelerate relaxation toward the magnetic ground state; tests with
smaller values produce the same final configurations but require longer
trajectories.

\begin{figure*}[htbp]
  \centering
  \includegraphics[width=\textwidth]{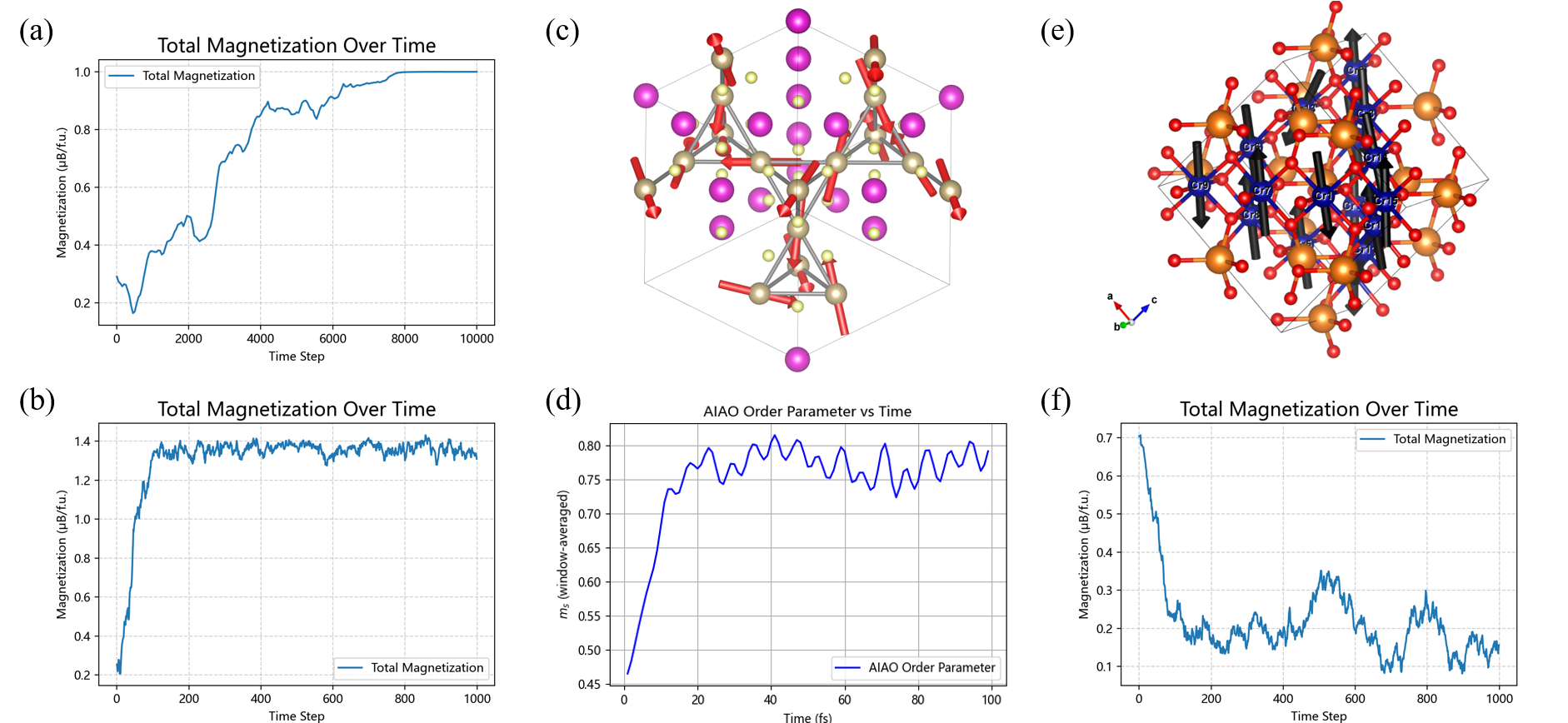}
  \caption{Magnetic ground-state validation across four representative systems.
           (a)~Monolayer CrI$_3$: convergence of the out-of-plane magnetization
           from a random initial state to the ferromagnetic ground state.
           (b)~Bcc Fe: convergence of the total magnetization to the collinear
           ferromagnetic state at 300~K.
           (c)~Cd$_2$Os$_2$O$_7$: final spin configuration showing the AIAO
           arrangement.
           (d)~Cd$_2$Os$_2$O$_7$: time evolution of the AIAO order parameter
           $m_s$ converging to $\approx 0.9$.
           (e)~MgCr$_2$O$_4$: spontaneous tetragonal distortion ($c/a$ ratio)
           accompanying the onset of collinear antiferromagnetic order during
           annealing from 50~K to 5~K.
           (f)~MgCr$_2$O$_4$: final spin configuration.}
  \label{fig:validation}
\end{figure*}

We first assess the method for two representative collinear ferromagnets,
monolayer CrI$_3$ and bcc Fe, which provide complementary tests in a
two-dimensional local-moment magnet and a three-dimensional itinerant metal,
respectively.
Monolayer CrI$_3$ is a prototypical two-dimensional van~der~Waals ferromagnet
with an out-of-plane Ising easy axis~\cite{Huang2017}; the computational details are summarized
in Table~\ref{tab:validation}.
Starting from fully random spin orientations on the Cr sites, the simulation
converges within 2~ps to the correct out-of-plane ferromagnetic ground state
[Fig.~\ref{fig:validation}(a)].
Bcc Fe, by contrast, is a canonical itinerant ferromagnet that still supports
well-defined local moments, thereby providing a useful test of the present
framework in a metallic system, with the corresponding simulation parameters
also listed in Table~\ref{tab:validation}.
When initialized from random moment orientations, the simulation at 300~K
develops the expected collinear ferromagnetic order, with the magnetization
fluctuating around a finite value corresponding to an average moment of
approximately $2\,\mu_\mathrm{B}$ per Fe atom [Fig.~\ref{fig:validation}(b)].
These two cases confirm that the present framework reproduces collinear
ferromagnetic order in both insulating and metallic systems.

Cd$_2$Os$_2$O$_7$ is a correlated 5$d$ oxide that exhibits a non-collinear
all-in-all-out (AIAO) spin configuration, in which the Os moments on each
tetrahedron either all point inward or all point
outward~\cite{Shinaoka2012}; computational details are given in
Table~\ref{tab:validation}.
Starting from random spin orientations on all Os sites, the simulation
converges rapidly to the AIAO arrangement.
To quantify the degree of ordering we employ the normalized AIAO order
parameter~\cite{Reimers1992,Zhang2021}
\begin{equation}
  m_s^2 = \frac{4}{N^2} \sum_{a=1}^{4}
    \left| \sum_i \mathbf{M}_i^a \right|^2 ,
  \label{eq:ms}
\end{equation}
where $\mathbf{M}_i^a$ is the magnetic moment of the $a$-th Os in the $i$-th
unit cell and $N$ is the total number of Os atoms.
The converged value $m_s \approx 0.9$, slightly below unity due to finite-size
effects in the single conventional unit cell, confirms that the simulation
reproduces the AIAO magnetic ground state
[Fig.~\ref{fig:validation}(c,d)].

To test the method in a system with strong geometric frustration, we simulate
the spinel oxide MgCr$_2$O$_4$~\cite{Xiang2011}, in which the magnetic
Cr$^{3+}$ ions occupy a pyrochlore sublattice with competing nearest-neighbor
antiferromagnetic exchange interactions; computational details are given in
Table~\ref{tab:validation}.
The SLD trajectory is annealed from 50~K to 5~K over 5~ps, starting from the
ideal cubic structure with random magnetic moments.
During the anneal the simulation captures a spontaneous symmetry lowering: the
lattice relaxes into an approximately tetragonal structure with
$c < \sqrt{2}\,a$, accompanying the establishment of collinear
antiferromagnetic order with propagation vector $\mathbf{q} = (0,0,0)$
[Fig.~\ref{fig:validation}(e,f)].
Both the structural distortion and the magnetic ground state are consistent
with the spin-driven Jahn--Teller transition to the $I4_1/amd$ phase predicted
by Xiang et al.~\cite{Xiang2011} and observed experimentally below 12.5~K~\cite{Chatterji2009},
demonstrating that the present framework can capture phenomena in which spin
and lattice order emerge cooperatively.

Taken together, these four systems---spanning a two-dimensional magnetic
insulator, an itinerant 3$d$ metal, a correlated 5$d$ oxide, and a
geometrically frustrated spinel---demonstrate that the present \textit{ab
initio} SLD framework can reproduce qualitatively distinct magnetic ground
states, including collinear ferromagnetic, non-collinear all-in-all-out, and
antiferromagnetic orders, without material-specific parameter adjustments.
In all cases, the simulations were initialized from fully random spin
configurations and required no prior assumption about the symmetry or type of
the target magnetic order, confirming the predictive, rather than postdictive,
character of the method.
Moreover, the spontaneous tetragonal distortion accompanying magnetic ordering
in MgCr$_2$O$_4$ exemplifies a class of phenomena where the spin and lattice
degrees of freedom are fundamentally coupled; such behavior cannot be captured
by atomistic spin-dynamics simulations in which the lattice is frozen,
highlighting the need for a coupled SLD treatment.

\begin{table}[htbp]
  \caption{Computational parameters for the four validation systems.}
  \label{tab:validation}
  \begin{ruledtabular}
  \begin{tabular}{lcccc}
    & CrI$_3$ & bcc Fe & Cd$_2$Os$_2$O$_7$ & MgCr$_2$O$_4$ \\
    \hline
    Supercell        & $2\times2\times1$ & $2\times2\times2$
                     & conv.\ cell & conv.\ cell \\
    Atoms            & 72  & 16  & 40  & 56 \\
    $k$-mesh         & $2\times2\times1$ & $4\times4\times4$
                     & $2\times2\times2$ & $3\times3\times3$ \\
    Cutoff (eV)      & 500 & 500 & 500 & 500 \\
    $T$ (K)          & 1   & 300 & 1   & 50$\to$5 \\
    $\Delta t$ (fs)  & 1   & 0.2 & 0.5 & 0.5 \\
    Duration (ps)    & 2   & 0.4 & 1   & 5 \\
    $\alpha$         & 0.2 & 0.2 & 0.2 & 0.2 \\
  \end{tabular}
  \end{ruledtabular}
\end{table}

\section{\label{sec:spinGNN}Training-Data Benchmark for SLD-Oriented SpinGNN Models}
For spin-aware machine-learning models, training-data construction is nontrivial
because the relevant configuration space spans both atomic and magnetic degrees
of freedom.
A common reference strategy is to combine structural snapshots with
independently randomized magnetic configurations.
Using SpinGNN~\cite{SpinGNN}, a spin-aware graph neural network model, as a
concrete test case, we compare models trained on randomized-spin reference data
with models trained on data generated from first-principles SLD trajectories.
To make this comparison interpretable, we distinguish between the effects of
training distribution, training-set size, wall-time budget, and temporal
correlation in trajectory-derived data.

For this benchmark, both training datasets are constructed from the same pool of
600 initial atomic structures of an 80-atom BiFeO$_3$ supercell with 16 Fe sites
carrying local moments.
These structures were sampled at equal intervals from a single $NpT$
molecular-dynamics trajectory generated with the VASP machine-learning force
field (MLFF) across 300--900~K.
For the reference dataset, each initial structure is combined with 10 fully
randomized spin configurations on the Fe sites, following the fully
randomized-spin strategy of Ref.~\cite{SpinGNN}.
Each resulting spin-lattice configuration is then evaluated by a separate static
constrained-moment DFT calculation.
For the SLD-derived dataset, each initial structure is used to initialize a
20-step first-principles SLD trajectory, and the 20 trajectory snapshots are
used as SLD-derived data points.
The two datasets therefore share the same pool of initial atomic structures but
differ in how the spin configurations are generated: fully randomized in the
reference case and co-evolved with the lattice along SLD trajectories in the
SLD-derived case.
To avoid information leakage from correlations among data derived from the same
initial structure, the training and validation sets are split at the level of
initial structures rather than individual data points.
Because the intended application is subsequent SLD simulation, model performance
is evaluated on an independent SLD trajectory that is excluded from both the
training and validation sets.

We benchmark models trained on SLD-derived data against those trained on the
randomized-spin reference dataset under three comparison protocols.
In the first, the SLD-derived and reference models are trained on datasets
containing exactly the same number of spin-lattice configurations.
In the second, the two models are trained on datasets generated under the same
wall-time budget, so they are compared at equal data-generation cost rather than
at a fixed number of training configurations.
Under the computational settings used here, generating one configuration for the
randomized-spin reference dataset requires approximately the same wall time as
generating two SLD trajectory snapshots.
This cost difference arises because consecutive SLD configurations are close in
phase space, so each electronic step can be initialized from the converged state
of the preceding one and typically requires fewer SCF iterations than an
independently initialized static reference calculation.
In the third protocol, we subsample the fixed-wall-time SLD dataset to match the
number of training configurations in the reference dataset.
This allows us to reduce temporal redundancy and assess the impact of
correlations between adjacent trajectory snapshots.

\begin{figure*}[htbp]
  \centering
  \includegraphics[width=\textwidth]{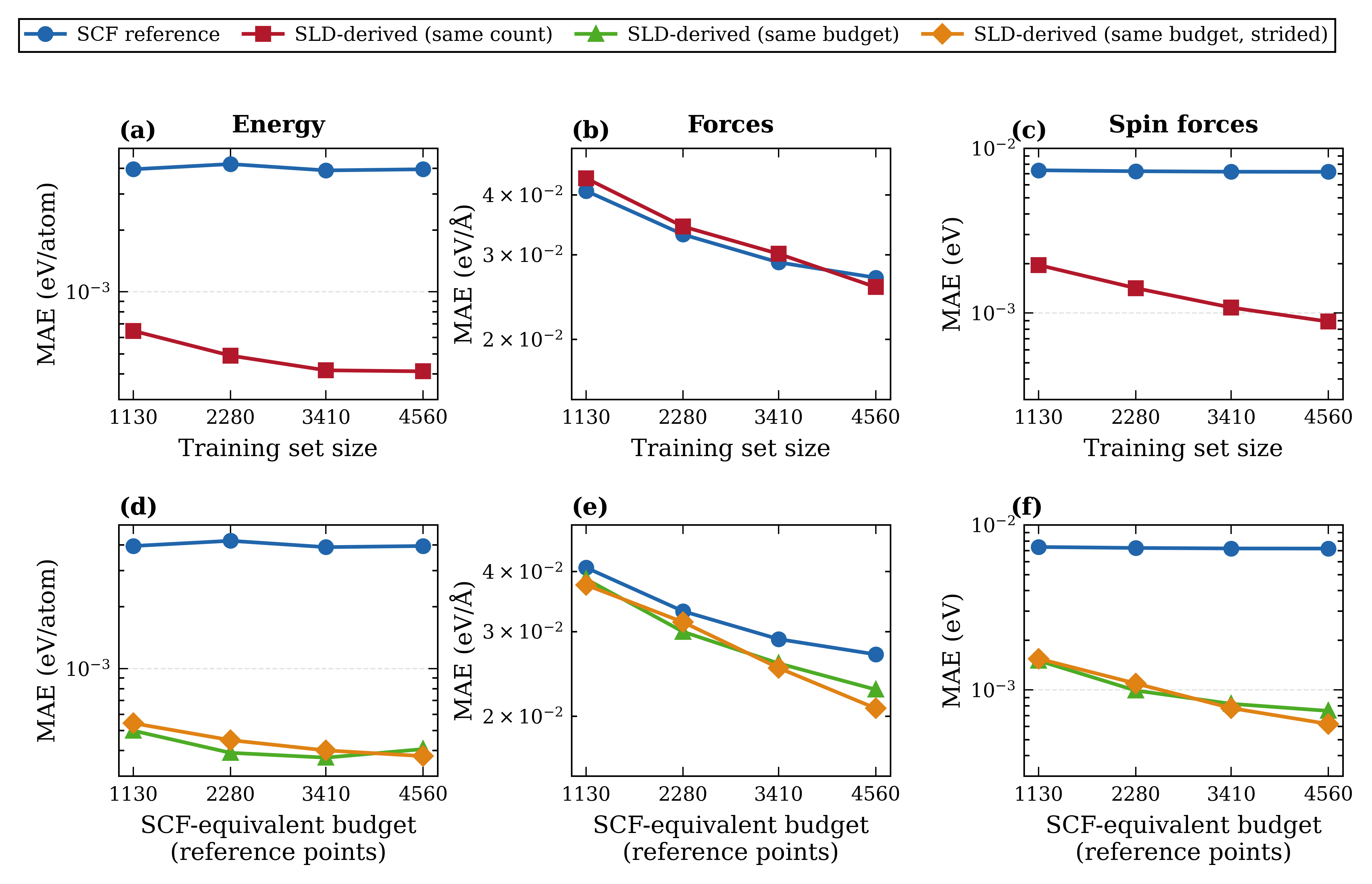}
  \caption{Training-data benchmark for SLD-oriented SpinGNN models in
           BiFeO$_3$. All mean absolute errors are evaluated on an independent
           SLD test trajectory. Panels (a)--(c) show the fixed-count comparison
           between the SCF reference and SLD-derived datasets for energy,
           forces, and spin forces, respectively. Panels (d)--(f) show the
           fixed-budget comparison among the SCF reference, the SLD-derived
           dataset generated at the same SCF-equivalent budget, and its strided
           counterpart used to reduce temporal correlation. The horizontal axis
           in the top row is the training-set size, whereas that in the bottom
           row is the SCF-equivalent data-generation budget, expressed in units
           of reference SCF-point cost. One SCF reference point requires
           approximately the same wall time as two saved SLD snapshots.}
  \label{fig:spingnn_benchmark}
\end{figure*}

Figures~\ref{fig:spingnn_benchmark}(a)--\ref{fig:spingnn_benchmark}(c) summarize
the benchmark results under the fixed-count protocol.
At fixed training-set size, the models trained on SLD-derived data yield
prediction errors on the independent SLD test trajectory that are comparable
for forces, yet reduced by approximately one order of magnitude for energy
and spin forces, relative to the randomized-spin reference models.
This suggests that the SLD-derived data provide a training distribution that is
better aligned with the independent SLD test trajectory, in the sense that it
yields lower prediction errors at the same training-set size.

Under a fixed wall-time budget, the models trained on SLD-derived data show
lower MAEs than the randomized-spin reference models on the independent SLD test
trajectory.
Combined with the fixed-count results, this shows that the stronger performance
of the SLD-derived models at equal wall-time arises from two factors: a more
effective training distribution and a larger number of training configurations
generated within the same computational budget.
When the fixed-wall-time SLD dataset is subsampled to reduce temporal
correlations at approximately matched training-set size, the resulting models
achieve the lowest MAEs across all three protocols, consistent with reduced
overfitting to redundant sequential configurations.
Taken together, this benchmark demonstrates that dynamically co-evolved
spin-lattice trajectories naturally provide a highly effective training
distribution for downstream SLD applications, as they directly sample the
coupled phase space explored during such dynamics.

\section{\label{sec:discussion}Discussion and Conclusions}

The \textit{ab initio} SLD framework presented in this work circumvents the reliance on pre-fitted
exchange interaction parameters typically required in atomistic spin-dynamics
simulations~\cite{Antropov1996,Skubic2008,Evans2014}.
By evaluating interatomic forces and effective magnetic fields directly from
constrained-moment DFT at each time step, the method enables the simultaneous
and mutually consistent evolution of both spin and lattice degrees of freedom.
Furthermore, while recent developments suggest that site-specific penalty
parameters can further refine constraint accuracy~\cite{Cai2023},
our tests confirm that a global $\lambda$ is sufficient for the dynamics
targeted here; the resulting trajectories remain highly insensitive to its
exact magnitude (see Supplemental Material).
The validation across four structurally and electronically distinct
systems---encompassing ferromagnetic, non-collinear, and geometrically
frustrated magnetic orders---demonstrates that no \textit{ad hoc} parameter
adjustments are required to reliably recover these diverse ground states from
random initial conditions.
The underlying approximations of spin adiabaticity and atomic-moment locality
restrict applicability to systems where longitudinal electronic relaxation is
sufficiently fast to justify treating moment magnitudes as instantaneously
adapting rather than independent dynamical variables.
This condition is well satisfied by a broad class of local-moment magnets and
by itinerant systems that still sustain well-defined local moments over the
time scales of interest, such as bcc Fe, but may break down in weak itinerant
magnets or mixed-valence systems where longitudinal fluctuations of the moment
magnitude are no longer negligible relative to the orientational
dynamics~\cite{Ruban2007,Stocks1998}.
Within this regime, while the Gilbert damping parameter $\alpha$ must be
assigned a physically realistic value for finite-temperature dynamical studies,
it serves merely to control the relaxation time scale when searching for the
ground state, leaving the converged magnetic order unaffected as confirmed by
our tests.

Beyond its utility for direct physical simulations, the BiFeO$_3$ benchmark
demonstrates that the generated SLD trajectories can serve as an effective
source of training data for spin-aware machine-learning potentials,
particularly when the downstream task is itself a spin-lattice dynamical
simulation.
We emphasize, however, that this conclusion does not establish SLD-derived
data as universally superior to all possible magnetic-data generation
strategies, nor does it imply that randomized magnetic configurations are
always an inappropriate baseline.
Rather, it demonstrates that for the present SpinGNN model and test protocol,
SLD-derived data provide a training distribution that is more representative
of the configurations actually encountered during independent SLD simulations
than a baseline constructed from identical structural snapshots with
randomized spins.
Furthermore, the matched-count and strided-sampling comparisons confirm that
the temporal correlation inherent to trajectory data does not outweigh this
advantage.

The physical origin of this distributional advantage lies in the fact that SLD
sampling naturally preserves the dynamic coupling between lattice distortions
and magnetic order, thereby avoiding the generation of statistically
improbable, high-energy configurations.
As either the simulated supercell size or the complexity of its magnetic
degrees of freedom increases, uniform random exploration of the
high-dimensional spin-configuration space becomes increasingly inefficient,
making SLD-based dynamical sampling particularly advantageous for generating
training data for spin-aware machine-learning models.
This framework also provides a natural route toward active-learning
workflows, in which model uncertainty along SLD trajectories triggers new
first-principles calculations in previously undersampled regions.

In summary, the present work establishes a practical fully \textit{ab initio}
framework for coupled spin-lattice dynamics in magnetic materials, in which
both interatomic forces and effective magnetic fields are obtained
self-consistently at each time step from constrained-moment DFT calculations.
The successful reproduction of diverse magnetic ground states from random
initial conditions demonstrates that this framework is applicable to a broad
class of magnets with robust local moments and provides a direct route to
finite-temperature simulations of spin-wave and magnon-phonon dynamics as well
as magnetostructural transitions.
The BiFeO$_3$/SpinGNN benchmark further shows that the resulting SLD
trajectories can furnish effective training data for spin-aware
machine-learning models for downstream spin-lattice simulations.
Together, these results establish a direct connection between first-principles
spin-lattice dynamics and data-driven modeling of magnetic materials.

\begin{acknowledgments}
We acknowledge financial support from the National Key R\&D Program of China
(Grant No.~2022YFA1402901), the National Natural Science Foundation of China
(NSFC, Grant No.~12188101), the Shanghai Science and Technology Program
(No.~23JC1400900), the Guangdong Major Project of Basic and Applied Basic Research
(Future Functional Materials under Extreme Conditions, Grant No.~2021B0301030005),
the Shanghai Pilot Program for Basic Research at Fudan University (No.~23TQ017),
the robotic AI-Scientist platform of the Chinese Academy of Sciences,
and the New Cornerstone Science Foundation.
\end{acknowledgments}

\bibliographystyle{apsrev4-2}
\bibliography{refs}

\end{document}